\documentclass[amsmath,amssymb,noshowpacs,aip,pop,
preprint,
citeautoscript  
]{revtex4-1}
\usepackage{graphicx}
\usepackage{dcolumn}
\usepackage{bm}

\usepackage[utf8]{inputenc}
\usepackage[T1]{fontenc}
\usepackage{mathptmx}
\usepackage{etoolbox}

\makeatletter
\def\@email#1#2{%
	\endgroup
	\patchcmd{\titleblock@produce}
	{\frontmatter@RRAPformat}
	{\frontmatter@RRAPformat{\produce@RRAP{*#1\href{mailto:#2}{#2}}}\frontmatter@RRAPformat}
	{}{}
}%
\makeatother

\newcommand\bb{\boldsymbol{b}}
\newcommand\bB{\boldsymbol{B}}
\newcommand\bR{\boldsymbol{R}}

\begin{document}
	
	\title[Piecewise field-aligned FEM]{Piecewise Field-Aligned Finite Element Method for Multi-Mode Nonlinear Particle Simulations in tokamak plasmas}
	\author{Zhixin Lu}
	\email{{zhixin.lu@ipp.mpg.de}}
	
	\author{Guo Meng}
	\author{Eric Sonnendrücker}
	\author{Roman Hatzky} 
	\affiliation{Max Planck Institute for Plasma Physics, 85748 Garching, Germany}
	
	\author{Alexey Mishchenko}
	\affiliation{Max Planck Institute for Plasma Physics, 17491 Greifswald, Germany}
	
	\author{Fulvio Zonca}
	\affiliation{Center for Nonlinear Plasma Science; C.R. ENEA Frascati, C.P. 65, 00044 Frascati, Italy}
	
	\author{Philipp Lauber}
	\affiliation{Max Planck Institute for Plasma Physics, 85748 Garching, Germany}
	
	\author{Matthias Hoelzl}
	\affiliation{Max Planck Institute for Plasma Physics, 85748 Garching, Germany}
	
	\date{\today}
	\begin{abstract}
		This paper presents a novel approach for simulating plasma instabilities in tokamak plasmas using the piecewise field-aligned finite element method in combination with the particle-in-cell method. 
		Our method traditionally aligns the computational grid but defines the basis functions in piecewise field-aligned coordinates to avoid grid deformation while naturally representing the field-aligned mode structures. 
		This scheme is formulated and implemented numerically. It also applied to the unstructured triangular meshes in principle. We have conducted linear benchmark tests, which agree well with previous results and traditional schemes.
		Furthermore, multiple-$n$ simulations are also carried out as a proof of principle, demonstrating the efficiency of this scheme in nonlinear turbulence simulations within the framework of the finite element method.
	\end{abstract}
	
	\maketitle
	
	\section{Introduction}
	\label{sec:introduction}
	A variety of waves and instabilities in tokamak plasmas are characterized by scale separation between the parallel and perpendicular directions to the magnetic field ${\bf B}$, namely, $k_{||}\ll k_\perp$, where $k_{||}$ and $k_\perp$ are the wave vectors in the parallel and perpendicular directions, respectively. To simulate these waves and instabilities, especially in the high $n$ limit, where $n$ is the toroidal mode number, theoretical and numerical schemes have been developed including the Ballooning representation \cite{connor1979high,cheng1985high}, the mode structure decomposition (MSD) approach \cite{zonca2014theory,lu2012theoretical,lu2013mixed,lu2017symmetry} and the flux-coordinate independent (FCI) scheme \cite{hariri2013flux,stegmeir2018grillix}. As a specific numerical treatment, the field-aligned coordinates and flux tube method have been developed for gyrofluid turbulence simulations \cite{beer1995field} and have shown to be an efficient scheme in gyrokinetic simulations \cite{lin1998turbulent,wang2011trapped}. 
	The metric procedure for ﬂux tube treatments of toroidal geometry has been developed previously to avoid grid deformation \cite{scott2001shifted}. 
	
	While the field-aligned coordinates used in most codes have been based on the finite difference scheme, their application in the framework of the finite element method (FEM) is still not available in particle simulations of magnetically confined plasmas even though many hybrid or gyrokinetic codes are based on FEM \cite{kleiber2024euterpe,lanti2020orb5,hatzky2019reduction,Huijsmans2023eps,hoelzl:hal-04403692,holderied2021mhd}. Previous studies have introduced the field-aligned discontinuous Galerkin method for anisotropic wave equations \cite{dingfelder2020locally}. A partially mesh-free approach  has also been proposed for representing anisotropic spatial variations along field lines \cite{mcmillan2017partially}.  FEM has advantages in terms of conservation properties and high accuracy in particle simulations. Gyrokinetic particle codes have been developed for the studies of turbulence, Alfv\'en waves, and energetic particle physics in both tokamaks \cite{lanti2020orb5,Huijsmans2023eps,lu2019development,hoelzl:hal-04403692} and stellarators \cite{jost2001global,kleiber2024euterpe}. While various schemes have been developed to efficiently represent the field-aligned mode structures such as the Fourier filter in ORB5 \cite{lanti2020orb5}, the phase factor in EUTERPE, and the high order differential operator-based spatial filter in JOREK \cite{Huijsmans2023eps}, a field-aligned FEM approach has not yet been developed. 
	
	In this work, we focus on the formulation and implementation of the field-aligned finite element method. The scheme we proposed is characterized by two key features: 
	\begin{enumerate}
		\item The computational grids are aligned in a traditional pattern without any shift.
		\item The finite element basis functions are defined on piecewise field-aligned coordinates, with each basis function being continuous along the magnetic field line. When the cubic spline is used, the $C^2$ continuity is maintained. 
	\end{enumerate}
	
	The rest of the paper is organized as follows. In Section \ref{sec:models}, the models and equations are detailed. The numerical schemes are described in Section \ref{sec:numeric}. In Section \ref{sec:results}, the simulation results are listed. In Section \ref{sec:conclusions}, we give the conclusions and outlook. 
	
	\section{Models and equations}
	\label{sec:models}
	\subsection{Formulation in 3D straight field line coordinates}
	\subsubsection{Coordinates and magnetic field}
	In the curvilinear coordinates $(X,Y,Z)$, consider the magnetic field pointing in the $(Y,Z)$ direction, the field is expressed as ${\bf B}=\nabla\psi_Y\times\nabla Z-\nabla\psi_Z\times\nabla Y$, where $\psi_Y$ and $\psi_Z$ are flux functions and are independent of $Y$ and $Z$, namely, $\psi_Y=\psi_Y(X)$, $\psi_Z=\psi_Z(X)$. The twisting of the magnetic field along $Y$ and $Z$ is described by 
	\begin{eqnarray}
	\bar{q}\equiv \frac{B^Z}{B^Y}=\frac{\partial\psi_Z}{\partial\psi_Y}\;\;,
	\end{eqnarray}
	where $\bar{q}=\bar{q}(X)$, $B^Y\equiv{\bf B}\cdot\nabla Y$, $B^Z\equiv{\bf B}\cdot\nabla Z$. Defining a Clebsch coordinate 
	\begin{eqnarray}
	A\equiv Y-\frac{Z}{\bar{q}} \;\;,
	\end{eqnarray}
	the magnetic field can be rewritten as
	\begin{eqnarray}
	{\bf B}=-\bar{q} (\nabla\psi_Y\times\nabla A)
	=-\bar{q}(\partial_X\psi_Y)\nabla X\times\nabla A \;\;.
	\end{eqnarray}
	Two Clebsch coordinates systems can be chosen as follows,
	\begin{enumerate}
		\item $(X,A\equiv Y-Z/\bar{q},Z)$, which is well defined if $\bar{q}\ne 0$;
		\item $(X,K\equiv \bar{q}Y-Z,Z)$, which is well defined if $\bar{q}\ne \infty$.
	\end{enumerate}
	Other choices are possible by choosing the third coordinate as $Y$, giving the $(X,Y,A)$ or $(X,Y,K)$ coordinates. In the following discussion, we adopt the first choice $(X,A,Z)$ since we consider the studies of the tokamak plasmas for which $\bar{q}\ne0$ but $\bar{q}=\infty$ at the so-called ``X'' point.  
	
	\begin{figure}
		\centering
		\includegraphics[width=0.4\textwidth]{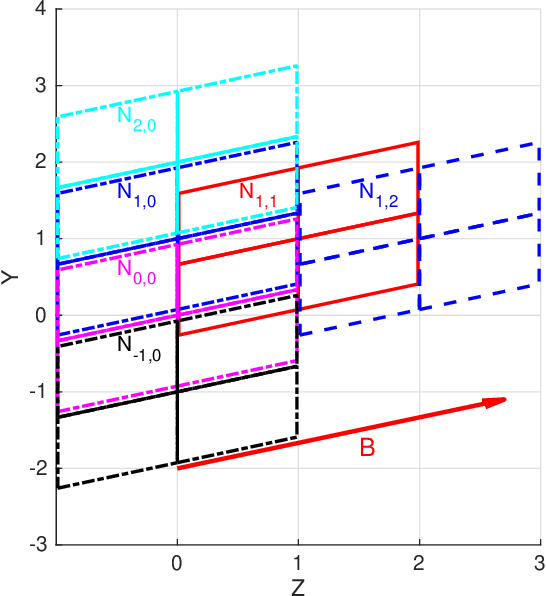}
		\caption{\label{fig:grids}The grids for simulations and the domains where the basis functions are defined. The dashed parallelograms, which match the color of basis function $N_{j,k}$, represent their domain. The overlaps between different basis functions are shown. The basis function $N_{j,k}$ overlaps with itself and two other functions $N_{j',k'}$ when $k'=k$, as shown by that $N_{0,0}$ overlaps with $N_{1,0}$ and $N_{-1,0}$. For $k'=k-1$, $N_{j,k}$ overlaps with the other four basis functions (the overlap between $N_{1,1}$ with $N_{j'=-1,0,1,2,k'=0}$). }
	\end{figure}
	
	\subsubsection{Piecewise field-aligned finite elements and partition of unity}
	The simulation grids are written as $(X_i,Y_j,Z_k)$, where $i,j,k$ are the grid indices along $X,Y,Z$ respectively. We show the construction of the piecewise field-aligned linear basis functions in Fig.~\ref{fig:grids}, where the grey lines indicate the $Y,Z$ grids. If cubic B-spline basis functions are adopted, each basis function covers four intervals in each direction. The red arrow indicates the magnetic field. Instead of constructing a global Clebsch coordinate $A$, the piecewise Clebsch coordinate $A_k$ depends on the index of the subdomain in $Z$ direction,
	\begin{eqnarray}
	A_k=Y-\frac{Z-Z_k}{\bar{q}}\;\;.
	\end{eqnarray}
	This definition is similar to the shifted metric procedure for the finite difference scheme \cite{scott2001shifted} but our method is adapted for the finite element method. 
	The piecewise field-aligned basis functions are defined in the coordinates $X,A_k,Z$. The function $\delta\phi$ in configuration space is represented as the superposition of the basis functions,
	\begin{eqnarray}
	\delta\phi=\sum_{i,j,k}\delta\phi_{i,j,k} N_i(X)N_j(A_k)N_k(Z)\;\;,
	\end{eqnarray}
	where $N$ is the basis function. In Fig.~\ref{fig:grids}, we show the linear basis functions in $(Y,Z)$ space for a given $X$, where $N_{j,k}\equiv N_j(A_k)N_k(Z)$. In the $Y$ direction, each basis function $N_{j,k}$ overlaps with itself and two other functions (in total three) $N_{j',k'}$ when $k'=k$, as shown by the overlap of $N_{0,0}$ with $N_{j'=1,-1;\;k'=0}$. However, when $k'\neq k$, each basis function $N_{j,k}$ overlaps with four other basis functions $N_{j',k'=k-1}$, as shown by the overlap between $N_{1,1}$ and $N_{j'=2,1,0,-1;\;k'=0}$. Additionally, $N_{j,k}$ overlaps with four other basis functions where $k'=k+1$; for instance, $N_{1,1}$ overlaps with $N_{j'=2,1,0,-1;\;k'=2}$.
	
	
	For each direction, the partition of unity is satisfied,
	\begin{eqnarray}
	\sum_i N_i(\alpha=\alpha_p)=1\;\;,\;\;\alpha\in[X,A,Z]\;\;,
	\end{eqnarray}
	where $\alpha_p$ indicates a given position of a marker or a sampling point. In the three-dimensional case,
	\begin{eqnarray}
	&&\sum_i\sum_j\sum_k N_i(X_p)N_j(A_p)N_k(Z_p) \nonumber\\
	&=&\sum_j\sum_k N_j(A_p)N_k(Z_p)\sum_iN_i(X_p) \nonumber\\
	&=&\sum_j\sum_k N_j(A_p)N_k(Z_p)  \nonumber\\ 
	&=&\sum_kN_k(Z_p)\sum_j N_j(A_p=Y_p-(Z_p-Z_k)/\bar{q}(X_p)) \nonumber\\
	&=&\sum_kN_k(Z_p)=1
	\end{eqnarray}
	We have demonstrated the partition of unity for this scheme, which has also been verified numerically in our implementation. An equivalent and intuitive proof of the partition of unity can also be found in a mesh-free scheme \cite{mcmillan2017partially}. 
	
	\subsection{Formulation in TRIMEG-GKX for tokamak plasmas}
	We have upgraded the previous version of the TRIMEG-GKX code \cite{lu2023full} to deal with experimental equilibria starting from the EQDSK file and to incorporate the piecewise field-aligned FEM. In a tokamak geometry with nested magnetic surfaces, the magnetic field is specified by 
	\begin{eqnarray}
	{\bf B}=\nabla\psi\times\nabla\phi+F\nabla\phi \;\;,
	\end{eqnarray}
	where $\psi$ is the poloidal magnetic flux function and $F$ is the poloidal current function.
	We formulate this method using tokamak coordinates $(r,\phi,\theta)$ since TRIMEG-GKX deals with the core plasma. This method can be readily extended to the whole device simulation using the unstructured triangular meshes in $(R,\phi,Z)$ coordinates, which enables the studies of the edge and scrape-off layer physics. A possible application is the combination of this piecewise field-aligned FEM and the unstructured triangular mesh in TRIMEG-C1  \cite{lu2019development,lu2024gyrokinetic}, which is beyond the scope of this work and will be reported in the future. 
	In tokamak coordinates, the radial-like coordinate is defined as $r=\sqrt{(\psi-\psi_{\rm edge})/(\psi_{\rm axis}-\psi_{\rm edge})}$, the poloidal-like coordinate $\theta=\arctan((Z-Z_0)/(R-R_0))$ for $\theta\in[-\pi/2,\pi/2)$. The Jacobian of $r,\phi,\theta$ is defined as $J=(\nabla r\times\nabla\phi\cdot\nabla\theta)^{-1}$. 
	
	The safety factor is
	\begin{eqnarray}
	q(r,\theta)\equiv\frac{{\bf B}\cdot\nabla\phi}{{\bf B}\cdot\nabla\theta} = \frac{JF}{R^2\partial_r\psi}\;\;.
	\end{eqnarray}
	The piecewise field-aligned coordinate $\eta_{k}$ in the subdomain centered at $\phi_k$ grid is obtained as follows,
	\begin{eqnarray}
	\label{eq:eta_integral}
	\eta_{k}(r,\theta,\phi)= \theta-\int_{\phi_k}^\phi {\rm d}\phi'\frac{1}{q(r,\theta',\phi')}  \;\;,
	\end{eqnarray}
	where the integral is along the magnetic field line and thus $q=q(r,\theta',\phi')$, $\phi_k$ and $\phi$ denote the integral's starting and end points, respectively. 
	Equation~\eqref{eq:eta_integral} is general and has been implemented in this work. Specifically, in the straight field line coordinates $r,\bar\theta,\bar\phi$, we readily get
	\begin{eqnarray}
	\label{eq:eta_integral_straightB}
	\eta_{k}(r,\bar\theta,\bar\phi)=\bar\theta-\frac{\bar\phi-\bar\phi_k}{\bar q} \;\;,
	\end{eqnarray}
	which can be constructed in other codes that use straight field line coordinates ($q$ is constant in a magnetic flux surface) such as EUTERPE \cite{kleiber2024euterpe}. 
	
\begin{figure}
		\centering
		\includegraphics[width=0.68\textwidth]{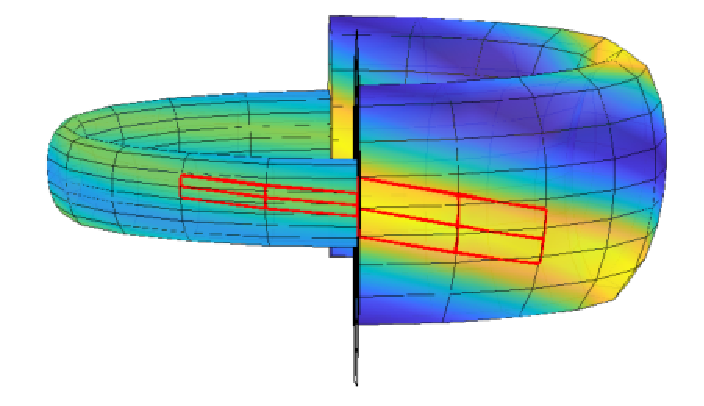}
		\caption{\label{fig:torus}Tokamak geometry with the piecewise field-aligned basis functions. The directions of the magnetic field and the alignment of the basis functions are different from the inner surface (left) to the outer surface (right). }
	\end{figure}

	The piecewise field-aligned FEM can be constructed for three-dimensional geometry with a similar treatment as that shown in Fig.~\ref{fig:grids} but with different directions of the magnetic field and a consequent different alignment of the basis functions at various radial locations, as indicated in Fig.~\ref{fig:torus}. 
	
	In tokamak plasmas, the pros and cons of the two choices of the field-aligned coordinates are shown in Tab.~\ref{tab:clebsch_choice}. Using the $r,\eta,\phi$ coordinates, it is possible to treat the `X' point at the separatrix of the tokamak plasma where $q=\infty$, which is one motivation of the TRIMEG code \cite{lu2019development}. By choosing $(r,\eta=\bar\theta-(\phi-\bar{\phi}_k)/\bar{q},\phi)$ instead of $(r,\bar\theta,\eta= \phi-\phi_k-\bar q \bar\theta)$, $\eta$ is more orthogonal to the other coordinate $\phi$ than to $\bar\theta$ since the magnetic field is mainly along the toroidal direction. 
	It should be noted that the equilibrium variables along $\phi$ with fixed $r,\eta=\bar\theta-(\phi-\phi_k)/\bar{q}$ are not constant anymore. The symmetry along the coordinate $\eta=\phi-\phi_k-\bar q \bar\theta$ remains if choosing $(r,\bar\theta,\eta=\phi-\phi_k-\bar q \bar\theta)$. Some detailed discussions can be found in the previous work and the references therein \cite{lu2012theoretical}.
	
	\begin{table}
		\centering
		\begin{tabular}{l c c}
			Choice of $\eta$                   & Pros & Cons \\ 
			$\eta=\bar\theta-(\phi-\phi_k)/\bar q,\phi$ & More orthogonal  & Symmetry breaking along $\phi$\\
			$\bar\theta,\eta=\phi-\phi_k-\bar q \bar\theta$        & Symmetry along $\eta$    & Less orthogonal  
		\end{tabular}
		\caption{\label{tab:clebsch_choice} Different choices of the Clebsch coordinates. For the sake of simplicity, we choose $\phi$ as the toroidal angle but adapt $\theta$ to $\bar\theta$ so that $(r,\bar\theta,\phi)$ is a straight field line coordinate system. }
	\end{table}
	
	\subsection{Discretization of the distribution function and marker initialization}
	Following the formulation in the previous $\delta f$ work \cite{hatzky2019reduction,lanti2020orb5,kleiber2024euterpe,mishchenko2019pullback}, $N_g$ markers are used with a given distribution,
	\begin{align}
	g(z,t)\approx \sum_{p=1}^{N_g} \frac{\delta[z_p-z_p(t)]}{J_z}\;\;,
	\end{align}
	where $z$ is the phase space coordinate, $N_g$ is the marker number, $\delta$ is the Dirac delta function, $J_z$ is the corresponding Jacobian, $z=({\bR},v_\|,\mu\equiv v_\perp^2/(2B))$ , and ${\bR}$ is the real space coordinate. For the full $f$ model, the total distribution of particles is represented by the markers,
	\begin{align}
	f(z,t)=C_{\rm{g2f}} P_{\rm{tot}}(z,t)g(z,t)
	\approx C_{\rm{g2f}} \sum_{p=1}^{N_g} p_{p,\rm{tot}}(t) \frac{\delta[z_p-z_p(t)]}{J_z}\;\;,
	\end{align}
	where the constant $C_{\rm{g2f}}\equiv N_f/N_g$, $N_{f/g}$ is the number of particles/markers, and $g$ and $f$ indicate the markers and physical particles respectively. 
	For each marker, 
	\begin{align}
	p_{p,\rm{tot}}(t)=\frac{1}{C_{\rm{g2f}}}\frac{f(z_p,t)}{g(z_p,t)}=\text{const} \;\;,
	\end{align}
	for collisionless plasmas since 
	\begin{align}
	\frac{{\rm d}g(z,t)}{{\rm d}t}=0 \;\;, \;\;
	\frac{{\rm d}f(z,t)}{{\rm d}t}=0 \;\;. 
	\end{align}
	The expression of $P_{\rm{tot}}(z,t)$ (and consequently, $p_{p,\rm{tot}}$) can be readily obtained 
	\begin{align}
	\label{eq:Ptot}
	P_{\rm{tot}}(z,t) = \frac{1}{C_{\rm{g2f}}}\frac{ f(z,t)}{g(z,t)}=\frac{n_f}{\langle n_f\rangle_V}\frac{\langle n_g \rangle_V}{n_g} \frac{f_v}{g_v} \;\;,
	\end{align}
	where  $n_f$ is the density profile and $f_v$ is the distribution in velocity space, namely, the particle distribution function $f=n_f({\bR})f_v(v_\parallel,\mu)$, $\langle \ldots\rangle_V$ indicates the volume average. There are different choices of the marker distribution functions as discussed previously \cite{hatzky2019reduction,lanti2020orb5}. 
	In our previous work \cite{lu2023full}, the markers are randomly distributed in the toroidal direction and in the $(R,Z)$ plane but the distribution in velocity space is identical to that of the physical particles, which leads to 
	\begin{align}\label{eq:ptot_general}
	p_{p,\rm{tot}}(z,t) =\frac{\phi_{\rm wid}SR}{V_{\rm tot}}\;\;,
	\end{align}
	where $\phi_{\rm wid}$ is the width of the simulation domain in the toroidal direction, $S$ is the area of the poloidal cross-section, $V_{\rm tot}$ is the total volume. Equation~\eqref{eq:ptot_general} is reduced to $p_{p,\rm{tot}}(z,t)={n_f}{R}/({\langle n_f\rangle_V}{R_0})$ for the tokamak equilibrium with concentric circular flux surfaces.
	In this work, as the first scheme, markers are loaded uniformly in $r^2$, $\theta$, and $\phi$ directions, which yields 
	\begin{align}
	\label{eq:load_scheme2}
	P_{\rm{tot}}(z,t) = \frac{n_f}{\langle n_f\rangle_V V}Jr_w\theta_w\phi_w\frac{r_{\rm mid}}{r} \;\;,
	\end{align}
	where  $r_{\rm mid}=(r_{\rm min}+r_{\rm max})/2$ is the middle location of the simulation domain,  $r_w$, $\theta_w$ and $\phi_w$ are the widths in the radial, poloidal and toroidal directions. 
	As the second scheme, markers are loaded uniformly in $r$, $\theta$ and $\phi$ directions, which yields
	\begin{align}
	\label{eq:load_scheme3}
	P_{\rm{tot}}(z,t) = \frac{n_f}{\langle n_f\rangle_V V}Jr_w\theta_w\phi_w \;\;.
	\end{align}
	In TRIMEG-GKX code, the scheme in Eq.~\eqref{eq:load_scheme2} is used more often since the poloidal Fourier filter is not used and thus we use the uniform loading in space to prevent a significant drop in the number of markers per cell near the plasma edge. 
	
	\subsection{Electrostatic gyrokinetic quasi-neutrality system}
	In this work, we adopt the electrostatic gyrokinetic model to minimize the technical complexity related to the electromagnetic model. The time step size is set small enough so that the so-called omega-H mode \cite{lee1987gyrokinetic} does not make the simulation crash. 
	
	\subsubsection{Gyro center's equations of motion}
	The gyro center's equations of motion are decomposed into the equilibrium part corresponding to that in the equilibrium magnetic field, and the perturbed part due to the perturbed field 
	\begin{eqnarray}
	\label{eq:gcmotion}
	{\boldsymbol{\dot R}} ={\boldsymbol{\dot R}}_0 + {\boldsymbol{\dot R}}_1 \;\;, \\
	\dot v_\| = \dot v_{\|,0}+ \dot{v} _{\|,1} \;\;.
	\end{eqnarray}  
	The gyro center's equations of motion are consistent with previous work \cite{mishchenko2019pullback,hatzky2019reduction,lanti2020orb5,mishchenko2023global,kleiber2024euterpe}, but are reduced to the electrostatic limit in this work. 
	\begin{eqnarray}
	\dot{\boldsymbol R}_0 
	&=& u_\| {\boldsymbol b}^* + \frac{m\mu}{qB^*_\|} {\boldsymbol b}\times\nabla B \;\;, 
	\\
	\dot v_{\|,0}
	&=& -\mu {\boldsymbol b}^*\cdot \nabla B \;\;,
	\\
	\dot{\boldsymbol R}_1
	&=& \frac{{\boldsymbol b}}{B^*_\|}\times \nabla \langle \delta\Phi\rangle \;\;, 
	\\
	\dot v_{\|,1}
	&=&  -\frac{q_s}{m_s} {\boldsymbol b}^*\cdot\nabla\langle\delta\Phi\rangle \;\;,  
	\end{eqnarray}
	where ${\boldsymbol b}^*={\boldsymbol b}+(m_s/q_s)v_\|\nabla\times{\boldsymbol b}/B_\|^*$, ${\boldsymbol b}={\boldsymbol B}/B$, $B_\|^*=B+(m_s/q_s)v_\|{\boldsymbol b}\cdot(\nabla\times{\boldsymbol b})$.
	
	The gyro centers' equations of motion are expressed in $(r,\phi,\theta)$ coordinates by keeping the dominant terms similar to the early treatment in the EUTERPE code \cite{jost2001global}.  The normalized equations are written where the velocity is normalized to $v_N$, $v_N=\sqrt{2T_N/m_N}$, $T_N$ is the temperature unit. The length is normalized to $R_N=1$ m and the time is normalized to $t_N=R_N/v_N$. In addition, we define the reference Larmor radius as $\rho_N=m_Nv_N/(eB_{\rm ref})$ since it appears with specific physics meaning related to the magnetic drift velocity and the finite Larmor radius effect, where $m_N$ is the proton mass, $B_{\rm ref}$ is the reference magnetic field. 
	The parallel velocity and the magnetic drift velocity in equilibrium are given by  
	\begin{eqnarray}
	\dot{r}_0&=&C_d\frac{F}{J}\partial_\theta B\;\;,\\
	\dot\theta_0&=&\frac{B^\theta}{B}v_\| -C_d\frac{F}{J}\partial_r B  \;\;,\\
	\dot\phi_0&=&\frac{B^\phi}{B}v_\| +C_d\partial_r\psi g^{rr}g^{\phi\phi}\partial_rB\;\;, \\
	C_d &=& \frac{\bar{m}_s}{\bar{e}_s}\rho_N\frac{B_{\rm ref}}{B^3}(v_\|^2+\mu B)
	\end{eqnarray}
	where $B^\alpha\equiv{\bB}\cdot\nabla\alpha$ is the contra-variant component of the equilibrium magnetic field,  $g^{\alpha\beta}\equiv\nabla\alpha\cdot\nabla\beta$ is the metric tensor. The equation due to the mirror force is given by
	\begin{eqnarray}
	\dot{v}_{\|,0}=-\frac{\mu}{JB}\partial_r\psi\partial_\theta B\;\;.
	\end{eqnarray}
	
	Regarding the perturbed equations of motion, the equilibrium variables are calculated in $(r,\phi,\theta)$ coordinates but $\partial_r\delta\phi$, $\partial_\theta\delta\phi$ and $\partial_\phi\delta\phi$ are calculated in $r,\phi,\eta$ coordinates. Especially, ${\bb}\cdot\nabla\delta\phi$ is directly calculated in $(r,\phi,\eta)$.
	
	The $E\times B$ velocity is given by 
	\begin{eqnarray}
	\dot{r}_1 &=& -\frac{\partial_r\psi}{B^2}\frac{g^{rr}}{R^2}\partial_\phi\delta\phi 
	+\frac{F}{JB^2}\partial_\theta \delta\phi \;\;, \\
	\dot\theta_1 &=& -\frac{\partial_r\psi}{B^2} \frac{g^{r\theta}}{R^2} \partial_\phi\delta\phi-\frac{F}{JB^2}\partial_r\delta\phi \;\;,\\
	\dot\phi_1&=&\frac{\partial_r\psi}{B^2R^2}
	\left(g^{rr}\partial_r\delta\phi+g^{r\theta}\partial_\theta\delta\phi\right)\;\;,
	\end{eqnarray}
	where $\partial_r\delta\phi$, $\partial_\theta\delta\phi$ and $\partial_\phi\delta\phi$ are in $(r,\phi,\theta)$ coordinates and need to be calculated from the Clebsch coordinates $r,\phi,\eta$. Using the chain rule, we readily have
	\begin{eqnarray}
	\partial_r|_{\phi,\theta} &=&\partial_r|_{\phi,\eta} +(\partial_r\eta)\partial_\eta\;\;,  \\
	\partial_\theta|_{r,\phi} &=& (\partial_\theta\eta)\partial_\eta  \;\;, \\
	\partial_\phi|_{r,\theta} &=&\partial_\phi|_{r,\eta} +(\partial_\phi\eta)\partial_\eta  \;\;.
	\end{eqnarray}
	
	The equation due to the parallel perturbed field is 
	\begin{eqnarray}
	\dot v_{\|,1}=-\frac{e_s}{m_s}\frac{B^\phi}{B}\partial_\phi|_{r,\eta}\delta\phi\;\;.
	\end{eqnarray}
	
	\subsection{Field equation}
	Since we focus on the electrostatic drift waves, the quasi-neutrality equation is adopted. In addition, the polarization density is expressed in the long wavelength limit. Then the quasi-neutrality equation is given by
	\begin{equation}
	\label{eq:poisson0}
	\nabla\cdot\left(
	G_{\rm P}\nabla_\perp\delta\phi
	\right)
	=-\sum_s e_s\delta n_s \;\;, \;\;G_P=\sum_s G_s\;\;,\;\;G_s=\frac{n_s}{\omega_{c,s} B} \;\;,
	\end{equation}
	where $\omega_{c,s}=|e_s| B/m_s$ is the cyclotron frequency.
	
	The rigorous expression in $(r,\theta,\phi)$ or $(r,\eta,\phi)$ is given by
	\begin{eqnarray}
	\frac{1}{J}\partial_\alpha (JG_s g^{\alpha\beta} \partial_\beta\delta\phi)
	-\frac{1}{J}\partial_\alpha(JG_s\nabla\alpha\cdot{\bb}{\bb}\cdot\nabla\beta\partial_\beta\delta\phi)
	=-\sum_s e_s\delta n_s \;\;, 
	\end{eqnarray}
	where $\alpha,\beta\in(r,\theta,\phi)$ for the previous scheme we have implemented \cite{lu2023full} or $\alpha,\beta\in(r,\eta,\phi)$ for the piecewise field-aligned FEM and the Einstein summation rule is adopted. Since the poloidal cross-section is a good approximation of the perpendicular surface, the simplified expression is given by
	\begin{eqnarray}
	\frac{1}{J}\partial_\gamma (JG_s g^{\gamma\iota} \partial_\iota\delta\phi)
	=-\sum_s e_s\delta n_s \;\;, 
	\end{eqnarray}
	where $\gamma,\iota\in(r,\theta)$ or $\gamma,\iota\in(r,\eta)$. In this work, both expressions are implemented numerically but no significant differences are found in the growth rate for the base case with $n=5$ in Section \ref{sec:results}. Thus, the following studies are all based on the simplified expression of the field equation. 
	
	When the piecewise field-aligned FEM is used, the metric tensor $g^{\alpha\beta}$ is calculated using the chain rule as follows (note $\eta=\eta_k$),
	\begin{eqnarray}
	g^{r\eta}&=&(\partial_r\eta) g^{rr} + (\partial_\theta\eta) g^{r\theta} + (\partial_\phi\eta) g^{\phi\theta} \;\;,   \\
	g^{\eta\eta}&=&(\partial_r\eta)^2 g^{rr} + (\partial_r\eta) (\partial_\theta\eta) g^{r\theta} + (\partial_\theta\eta) (\partial_r\eta)  g^{\theta r} +(\partial_\theta\eta)^2 g^{\theta\theta} \;\;,   \\
	g^{\phi\eta}&=&(\partial_r\eta) g^{r\phi}  + (\partial_\theta\eta) g^{\phi\theta} + (\partial_\phi\eta) g^{\phi\phi} \;\;.   
	\end{eqnarray}
	The metric tensors $g^{rr}$, $g^{r\phi}$ and $g^{\phi r}$ in $(r,\eta,\phi)$ are the same as those in $(r,\theta,\phi)$.
	
	\section{Numerical Implementation}
	\label{sec:numeric}
	
	\subsection{General description}
	The TRIMEG-GKX code is based on structured meshes for studying the core plasmas in tokamaks \cite{lu2021development,lu2023full}. It is written in Fortran. Object Oriented Programming (OOP) concepts are considered with a similar structure to the TRIMEG-C0/C1 code based on the unstructured meshes \cite{lu2019development,lu2024gyrokinetic}. The gyrokinetic field-particle system is decomposed into different classes, namely, equilibrium, particle, field, solver, and B-spline classes. The application of the gyrokinetic field-particle classes is constructed by other basic classes. The kernel of the Fortran code is about 14000 lines. The PETSc library is adopted to solve linear field equations using the KSP solver. The shared memory in the MPI3 standard is used to store the 3D field with affordable memory consumption. The equilibrium variables are represented using the B-splines \cite{williams}. The FEM is implemented using cubic splines with the details provided in our previous work \cite{lu2023full}.
	
	\subsection{The 3D field-aligned FEM solver and the mixed 2D1F solver}
	
	For the comparison with the 3D field-aligned FEM solver, a mixed 2D1F solver is also developed in this work, following the previous work \cite{lu2019development}, but using structured meshes and cubic splines. 
	For the 3D solver, the size of the grids is $(N_r,N_\theta,N_\phi)$ and $(N_{r,\rm{FEM}},N_{\theta,\rm{FEM}},N_{\phi,\rm{FEM}})$ basis functions are adopted to represent functions in the simulation domain, where $N_{r,\rm{FEM}}=N_r+\Delta N$, $N_{\theta,\rm{FEM}}=N_\theta$, $N_{\phi,\rm{FEM}}=N_\phi$, which are consistent with the boundary conditions, where $\Delta N=2$ since cubic splines are adopted. We apply the periodic boundary conditions in the $(\theta, \phi)$ directions and implement the Dirichlet and Neumann boundary conditions in the $r$ direction. 
	In poloidal and toroidal directions, the cubic finite element basis functions $N(x)$ are as follows
	\begin{equation}
	N_{\rm{cubic}}^{\rm inner}(x) =
	\begin{cases}
	4/3+2x+x^2+x^3/6 \;\;, & \text{if $x\in[-2,-1)$}\\
	2/3-x^2-x^3/2 \;\;,    & \text{if $x\in[-1,0) $}\\
	2/3-x^2+x^3/2 \;\;,    & \text{if $x\in[0,1) $} \\
	4/3-2x+x^2-x^3/6 \;\;, & \text{if $x\in[1,2) $} \;\;.
	\end{cases}       
	\end{equation}
	Along $\theta$ and $\phi$, the $i$th basis function is $N_i=N_{\rm{cubic}}(x+1-i)$. 
	In radial direction, $N_i$ is the same as those in poloidal/toroidal directions as $i\ge4$ or $i\le N_{r,\rm{FEM}}-3$. The first basis function is 
	\begin{equation}
	N_{\rm{cubic}}^{{\rm outer},1}(x) =
	\begin{cases}
	0 \;\;, & \text{if $x\in[-2,1)$}\\
	-x^3+6x^2-12x+8 \;\; & \text{if $x\in[1,2) $} \;\;.
	\end{cases}       
	\end{equation}
	The second basis function is
	\begin{equation}
	N_{\rm{cubic}}^{{\rm outer},2}(x) =
	\begin{cases}
	0 \;\;, & \text{if $x\in[-2,0)$}\\
	7x^3/6-3x^2+2x \;\;,    & \text{if $x\in[0,1) $} \\
	4/3-2x+x^2-x^3/6 \;\;, & \text{if $x\in[1,2) $} \;\;.
	\end{cases}       
	\end{equation}
	The third basis function is
	\begin{equation}
	N_{\rm{cubic}}^{{\rm outer},3}(x) =
	\begin{cases}
	0 \;\;, & \text{if $x\in[-2,-1)$}\\
	-x^3/3-x^2+2/3 \;\;,    & \text{if $x\in[-1,0) $}\\
	x^3/2-x^2+2/3 \;\;,    & \text{if $x\in[0,1) $} \\
	-x^3/6+x^2-2x+4/3 \;\;, & \text{if $x\in[1,2) $} \;\;.
	\end{cases}       
	\end{equation}
	The last three basis functions are symmetric mapping of the first three basis functions to the middle point of the simulation domain.  All radial basis functions are constructed according to  $N_i=N_{\rm{cubic}}(x+1-i)$, where $i\in[1,N_{r,\rm{FEM}}]$. 
	
	The weak form of the quasi-neutrality equations is 
	\begin{eqnarray}
	\label{eq:mat_poisson}
	\bar{\bar{M}}_{\mathrm{P},L,ii',jj',kk'} \cdot\delta\Phi_{i'j'k'}
	&&= C_{\mathrm P} \delta N^{i,j,k} \;\;,
	\end{eqnarray}
	where $\delta\Phi$ is normalized to $m_Nv_N^2/e$, and $C_P=1/\rho_N^2$. 
	
	For the 3D field-aligned FEM solver, 
	\begin{eqnarray}
	\bar{\bar{M}}_{\mathrm{P},L,ii',jj',kk'}
	&&=-\sum n_{0s} \bar{m}_s \frac{B^2_{\rm{ref}}}{B^2}\int {\rm d} r\, {\rm d}\theta\, {\rm d}\phi\, J \nabla_\perp\Tilde{N}_{ijk} 
	\cdot\nabla_\perp \Tilde{N}_{i'j'k'}\;\;,
	\nonumber \\ 
	\delta N^{i,j,k}
	&&=
	\sum_s C_{{\rm p2g},s}\sum_{p=1}^{N_g} w_p \Tilde{N}_{ijk}(r_p,\eta_{p,k},\phi_p) \;\;,\nonumber
	\end{eqnarray}
	where $\Tilde{N}_{ijk}(r,\eta,\phi)=N_i(r)N_j(\eta_k)N_k(\phi)$, $C_{{\rm p2g},s}=-\Bar{q}_s \langle n\rangle_V V_{\rm tot}/N_g$, $ \langle\ldots\rangle_V$ indicates the volume average, and $V_{\rm tot}$ is the total volume. 
	
	For the 2D1F solver, we have
	\begin{eqnarray}
	\bar{\bar{M}}_{\mathrm{P},L,ii',jj',kk'}
	&&=-\sum n_{0s} \bar{m}_s \frac{B^2_{\rm{ref}}}{B^2}\int {\rm d} r\, {\rm d}\theta\, {\rm d}\phi \, J \nabla_\perp[N_i N_j {\rm e}^{{\text i}k\phi}] 
	\cdot\nabla_\perp [N_{i'}N_{j'}{\rm e}^{-{\text i}k\phi}] \delta_{k,-k'} \;\;,
	\nonumber \\ 
	\delta N^{i,j,k}
	&&=
	\sum_s C_{{\rm p2g},s}\sum_{p=1}^{N_g} w_p N_i(r_p)N_j(\theta_p) {\rm e}^{-{\text i}k\phi_p} \;\;,\nonumber
	\end{eqnarray}
	where $N_i=N_i(r)$, $N_j=N_j(\theta)$, $\delta_{i,j}=1$ if $i=j$ and $\delta_{i,j}=0$ if $i\ne j$.
	
	\subsection{Matrix construction using the piecewise field-aligned FEM}
	There are different ways to calculate the matrix and stiffness matrices using the piecewise field-aligned FEM. Note that the partition of unity is always satisfied. The main difference is the numerical efficiency and accuracy in the matrix generation. 
	\subsubsection{The rigorous numerical integral}
	\label{subsubsec:rigorous_integral}
	The general form of the element of the mass/stiffness matrix is
	\begin{eqnarray}
	M_{ii',jj',kk'}&=&\int {\rm d}\Vec{x}\,
	\partial_{d_1}N_i(X)\partial_{d'_1}N_{i'}(X) \nonumber \\
	&\times& \partial_{d_2}N_j(A_k)\partial_{d'_2}N_{j'}(A_{k'})
	\partial_{d_3}N_k(Z)\partial_{d'_3}N_{k'}(Z) \;\;,
	\end{eqnarray}
	where $i,j,k$ indicate the row indices, $d_1$, $d_2$ and $d_3$ indicate the differential orders in the three directions, and $'$ indicates the variables of the matrix column. 
	We use the cubic B-spline basis functions. Each basis function $N_i$ is defined in four sections of the meshes. In $X$ and $Z$ directions, the overlaps of two basis functions can be one, two, three, or four sections. However, in $Y$ direction, the overlap can be a fractional length of the sections, as shown in Fig.~\ref{fig:grids}. As a result, the integral region needs to be identified first, as shown in Fig.~\ref{fig:gauss_integral}, where $\partial_{d_2}N_j(A_k)\partial_{d'_2}N_{j'}(A_{k'})$ needs to be integrated between the vertical red dashed line and the vertical blue dashed line. The Gauss-Legendre quadrature points are generated in this overlapping region and the integral is numerically calculated from the summation. This scheme is challenging to extend to unstructured triangular meshes since it is more complicated to identify the overlapping regions. 
	
	\subsubsection{The mixed-particle-wise-basis-wise construction}
\begin{figure}
		\centering
		\includegraphics[width=0.48\textwidth]{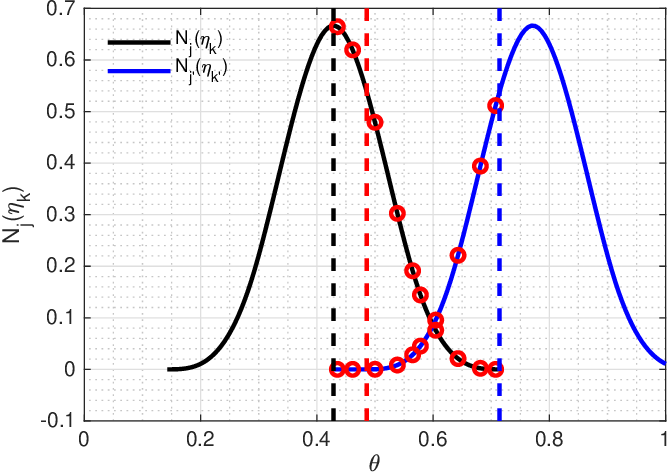}
		\caption{The numerical integral using the Gauss Legendre quadrature when $k'\ne k$ where $k$ and $k'$ are the indices in the toroidal direction. }
		\label{fig:gauss_integral}
	\end{figure}

	In addition to the rigorous numerical integral in Section~\ref{subsubsec:rigorous_integral}, the mixed-particle-wise-basis-wise construction is also possible with a minor loss of accuracy. As shown in Fig.~\ref{fig:gauss_integral}, the Gauss Legendre points are generated along the basis function of the row in the matrix, and the corresponding values of the basis function of the column in the matrix are calculated at the same $Y$ locations. Then the integral of the term $\partial_{d_2}N_j(A_k)\partial_{d'_2}N_{j'}(A_{k'})$ is calculated using the values at these points, indicated by the red circles in Fig.~\ref{fig:gauss_integral}. This scheme can be less accurate than the rigorous integral but is easier to extend to unstructured triangular meshes in TRIMEG-C0/C1 \cite{lu2019development,lu2024gyrokinetic}, as will be developed in the future.
	
	\subsubsection{Particle-wise construction using the Monte-Carlo integration and importance sampling}
	The mass and stiffness matrices are calculated using the Monte-Carlo integration, instead of the Gaussian quadrature, to avoid the complexities of identifying the nonconforming intersection surfaces of different volumes. 
	The integral can be generally calculated using the Monte-Carlo integration in connection with the importance sampling,
	\begin{eqnarray}
	I\equiv\int {\rm d}\vec{x} \, K(\vec{x})\approx
	\sum_i\frac{K(\vec{x})}{g(\vec{x})}\;\;,
	\end{eqnarray}
	where $K$ is the integrand, $g(\vec{x})$ is the distribution of the sampling points. For uniformly distributed sampling points, 
	\begin{eqnarray}
	I\approx
	\frac{V}{N}\sum_iK(\vec{x})\;\;.
	\end{eqnarray}
	In general geometry, uniform loading can be adopted along three coordinates $(X,Y,Z)$ and thus, \begin{eqnarray}
	g(\vec{x})=\frac{N}{JX_wY_wZ_w}\;\;,
	\end{eqnarray}
	where $J$ is the Jacobian, $X_w$, $Y_w$, and $Z_w$ are the width along $X$, $Y$ and $Z$ respectively. 
	This scheme can be less accurate than the rigorous integral and the mixed scheme but is easier to extend to two-dimensional unstructured triangular meshes or three-dimensional unstructured meshes.
	
	The properties of the three methods of matrix construction are listed in Tab.~\ref{tab:matrix_method}. While the rigorous scheme gives high accuracy, it can not be directly applied to the unstructured meshes adopted in TRIMEG-C1 since it is complicated to identify the overlapping regions of different triangles. The mixed and the Monte-Carlo schemes apply to the unstructured meshes but are not as accurate as the rigorous scheme. 
	
	\begin{table}
		\centering
		\begin{tabular}{l c c}
			Scheme                   & Pros & Cons \\ 
			Rigorous  & High accuracy & Not applicable for unstructured meshes\\
			Mixed         & Applicable for unstructured meshes & Less accurate than rigorous integral \\
			Monte-Carlo     & Applicable for unstructured meshes   &  Marker noise  ($1/\sqrt{N_g}$) 
		\end{tabular}
		\caption{\label{tab:matrix_method} The properties of different schemes for matrix construction.}
	\end{table}
	

	

	
	\section{Simulation results}
	\label{sec:results}
	
	In this work, the GA-STD case (core) parameters are adopted as reported in the previous benchmark work \cite{gorler2016intercode}. Concentric circular magnetic flux surfaces are adopted. The equilibrium density and temperature profiles denoted as $H(r)$, and the normalized logarithmic gradients $R_0/L_\mathrm{H}$, are given by 
	\begin{eqnarray}
	\label{eq:nTprofile}
	\frac{H(r)}{H(r_0)}  =\exp\left[-\kappa_\mathrm{H}w_\mathrm{H}\frac{a}{R_0}\tanh\left(\frac{r-r_0}{w_\mathrm{H}a}\right)\right]\;\;,  
	\end{eqnarray}
	where $L_\mathrm{H}=-\left[{\rm d}\ln H(r)/{\rm d}r\right]^{-1}$ is the characteristic length of profile $H(r)$ and $r_0=a/2$. 
	The ion-to-electron mass ratio is $m_{\rm i}/m_{\rm e}=1836$ and the deuterium ($m_{\rm i}/m_{\rm p}=2$) is the only ion species where $m_{\rm p}$ is the mass of a proton.
	The on-axis magnetic field $B_0=2\rm{T}$, $\rho^*=\rho_{\rm i}/a=1/180$, $\rho_{\rm i}=\sqrt{2T_{\rm i}m_{\rm i}}/(eB_0)$, aspect ratio $\epsilon=a/R_0=0.36$, $T_{\rm e}/T_{\rm i}=1$, characteristic length of temperature and density profiles $R_0/L_{T_{\rm i}}=-(d\rm{ln }T_{\rm i}/{\rm d}r)^{-1}=6.96$, $R_0/L_{T_{\rm e}}=-(\rm{d}\ln T_{\rm e}/\rm{d}r)^{-1}=6.96$, $R_0/L_{n}=-({\rm d}\ln n/{\rm d}r)^{-1}=2.23$, and collision frequency $\nu_{\rm coll}=0$.
	
	\subsection{Single toroidal harmonic simulations for benchmark}
\begin{figure}
		\centering
		\includegraphics[width=0.44\textwidth]{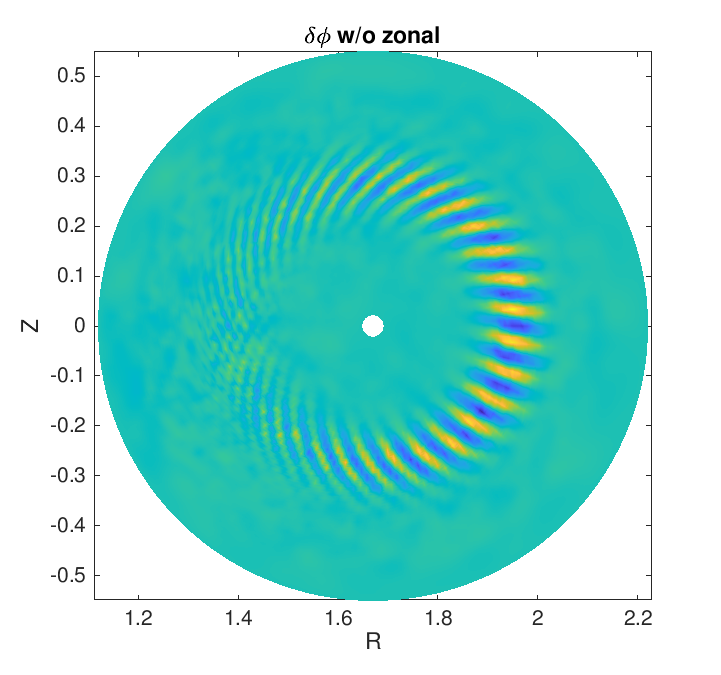}
		\includegraphics[width=0.44\textwidth]{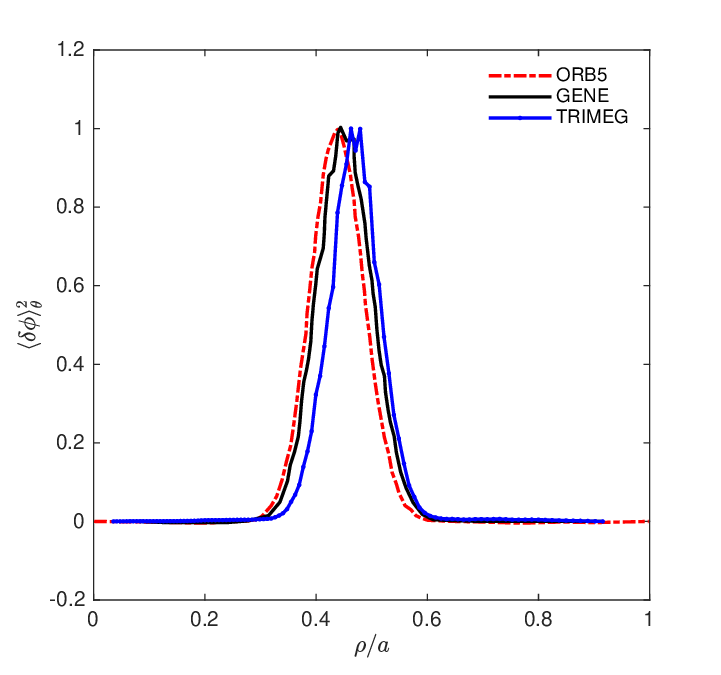}
		\caption{The two-dimensional mode structure (left) and the radial structure (right) of the ITG mode for $n=25$ }
		\label{fig:mode2d}
	\end{figure}

	The 2D1F field solver is used in the benchmark with the GENE results in the previous work \cite{gorler2016intercode}. 
	The simulation for the $n=25$ mode is carried out using $8\times10^6$ electrons and $10^6$~ions. For electrons, the gyro-average is switched off while the 4-point gyro-average is adopted for ions. The mass ratio is $m_{\rm i}/m_{\rm e}=1836$. The reference Larmor radius is $\rho_N=0.0033422$ m.  The time step size is $\Delta t=10^{-4} R_N/v_N$ for $n=25$. The 2D mode structure is shown in Fig.~\ref{fig:mode2d}. 
	
\begin{figure}
		\centering
		\includegraphics[width=0.48\textwidth]{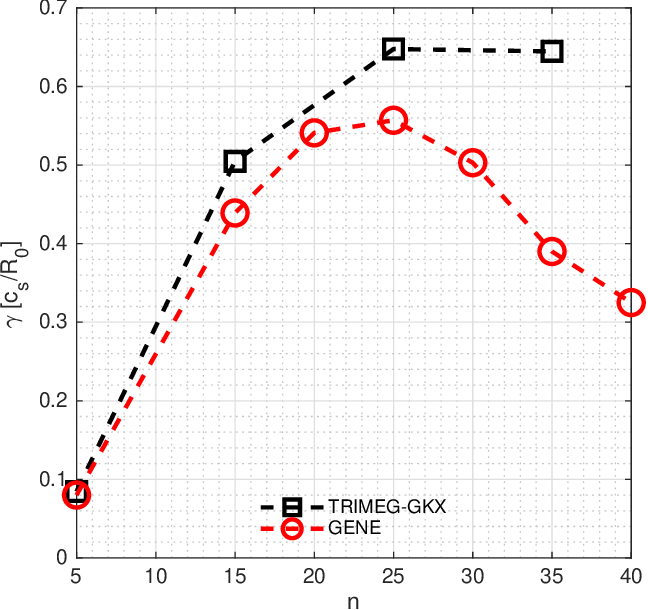}
		\caption{Growth rate of ITG mode for different values of the toroidal mode number $n$.}
		\label{fig:cbc_nominal_singlen}
	\end{figure}

	The growth rate of the ITG mode for different values of the toroidal mode number $n$ is studied. The growth rate is measured using the total field energy.  The total field energy is defined as 
	\begin{eqnarray}
	E_\Phi=-C_{\rm P}\int {\rm d}V \frac{1}{G_{\rm P}}\delta\bar\Phi\delta \bar N \;\;, \;\;
	\end{eqnarray}
	where $G_{\rm P}$ is defined in Eq.~\eqref{eq:poisson0} and $E_{\rm \Phi}$ is an approximate value of the field energy in $\delta\Phi$ in the limit $|k_\|/k_\perp|\ll 1$  and  $|\nabla_\perp\ln G_{\rm P}|/|\nabla\ln\delta\bar\Phi|\ll1$.
	A reasonably good agreement is observed between the results from the TRIEMG-GKX code and the GENE code, as shown in Fig. \ref{fig:cbc_nominal_singlen}. The time step size is small enough to  stabilize the omega-H mode and is at least smaller than $10^{-3}$ of the ITG period, sufficient to resolve the ITG mode accurately. The simulations with $4\times10^6$ and $8\times10^6$ electron markers give similar growth rates as an indicator of the convergence concerning the marker number. Note that the quasi-neutrality equation in the long wavelength is adopted in this work while in GENE, there is no truncation in the quasi-neutrality equation, which can lead to the discrepancy for high $n$ modes as shown in Fig. \ref{fig:cbc_nominal_singlen}.  The discrepancy at $n\approx32$ (namely, $k_y\rho_s\approx0.5$ using the notation in the previous work \cite{gorler2016intercode}) is expected. 
	
	\subsection{Comparison of the field-aligned FEM solver and the 2D1F solver}
	
\begin{figure}
		\centering
		\includegraphics[width=0.48\textwidth]{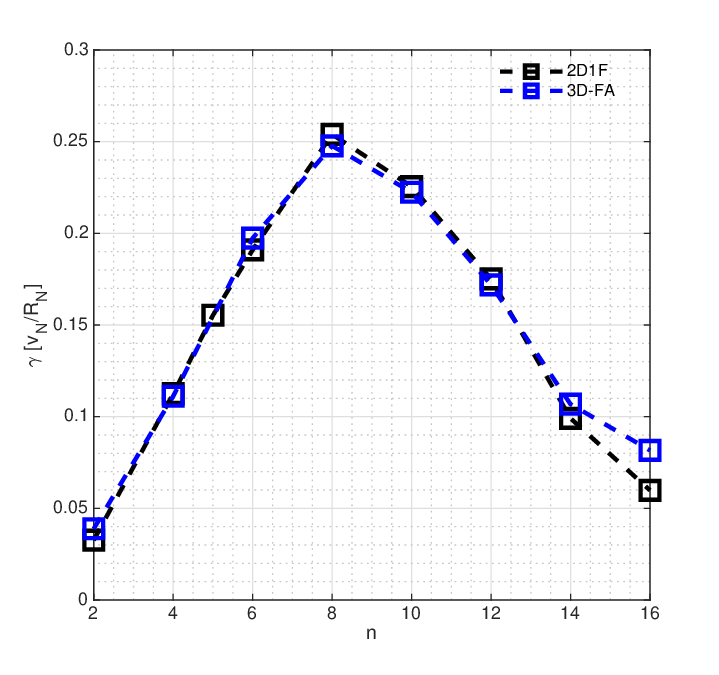}
		\caption{Growth rate and frequency of ITG mode for different values of the toroidal mode number $n$, using the 2D1F scheme and the 3D field-aligned FEM.}
		\label{fig:cbc_compare_2d1f_3d}
	\end{figure}

	The 3D piecewise field-aligned FEM solver is verified by comparing it with the 2D1F solver using the economical CBC parameters. Since the grids are still arranged in a traditional pattern without any shift, the toroidal Fourier filter can be readily applied in the 3D solver. 
	The parameters are the same as the nominal ones except the mass ratio $m_{\rm i}/m_{\rm e}=100$ and the reference Larmor radius $\rho_N=0.01$ m.
	From our simulations using various values of $\rho_{\rm N}=0.02, 0.01, 0.005$, the computational cost to avoid a numerical crash depends on several physics parameters as follows,
	\begin{eqnarray}
	C_{\rm comp}\propto 
	C_{\Delta t}C_{\rm marker} C_{m_{\rm e}} 
	\approx \rho_{\rm N}^{-1} \rho_{\rm N}^{-2} m_{\rm e}^{-1/2} \;\;,    
	\end{eqnarray}
	where $C_{\Delta t}$ is due to the reduction of the time step size $\Delta t$ as $\rho_{\rm N}$ decreases, $C_{\rm marker}$ is due to the increment of the marker number as $\rho_{\rm N}$ decreases, $C_{m_{\rm e}}$ is due to the reduction of the time step size $\Delta t$ as $m_{\rm e}$ decreases. 
	As a result, larger $\rho_N$ is adopted to make the simulation less costly. 
	For the 3D field-aligned FEM solver, the toroidal Fourier filter is applied to simulate a single toroidal harmonic. The comparison is shown in Fig. \ref{fig:cbc_compare_2d1f_3d}. A good agreement is observed.

	\subsection{Multi-toroidal harmonic simulations}
	
	The multi-$n$ nonlinear simulations are carried out without any Fourier filter. Buffer regions are applied in the inner and outer boundaries to minimize the noise near the simulation boundary. The nominal parameters are adopted except $m_{\rm i}/m_{\rm e}=100$. The grid numbers along the radial, poloidal and toroidal (parallel) directions are $N_r=96, N_\theta=192, N_\phi=8$, respectively. The time step size is $\Delta t=0.0025 R_N/v_N$. $32\times10^6$ electrons and $4\times10^6$ ions are used. 
	The simulation is run on 16 nodes (AMD EPYC Genoa 9554) of the Viper supercomputer at MPCDF, with 128 CPU cores on each node,  with a processor base frequency of $3.1$ GHz and a max turbo frequency of $3.75$ GHz. It takes $\sim 1.136$ hours to simulate one normalized time unit $t_{\rm N}=R_{\rm N}/v_{\rm N}$. 
	
\begin{figure}
		\centering
		\includegraphics[width=0.48\textwidth]{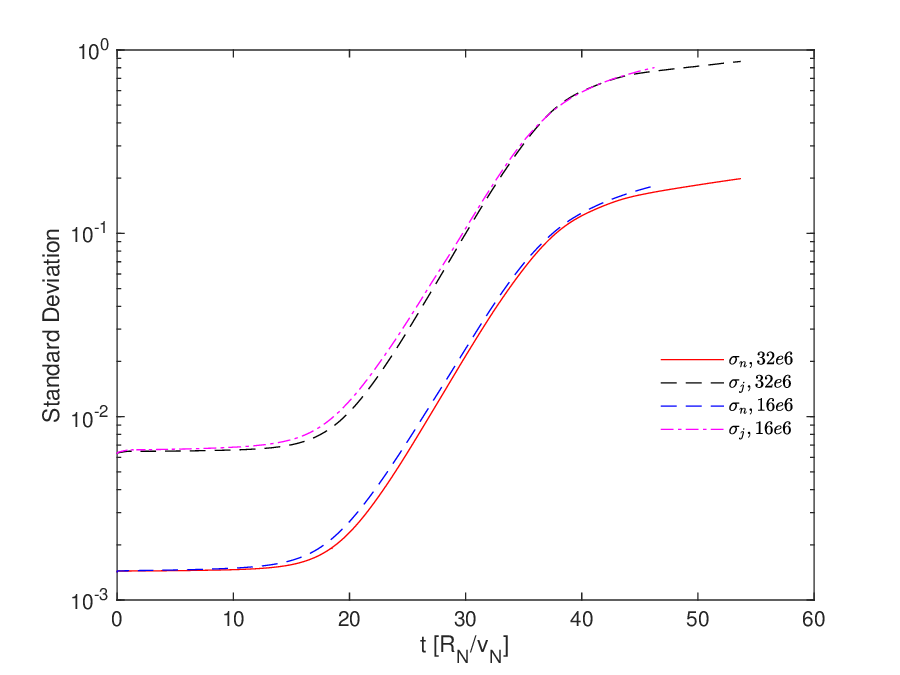}
		\caption{The standard deviation of the total perturbed density $\sigma_{\delta n}$ (red line) and current $\sigma_{\delta j}$ (black dashed line) ($32\times10^6$ electrons and $4\times10^6$ ions). A case with halved marker numbers ($16\times10^6$ electrons and $2\times10^6$ ions) is also shown where $\sigma_{\delta n}$ is given by the blue dashed line and  $\sigma_{\delta j}$ is given by the dashed magenta line. }
		\label{fig:noise1d}
	\end{figure}

	The statistic error is indicated by the unbiased estimator of the variance \cite{hatzky2019reduction}. Generally, the statistical error for $N_g$ markers is given as
	\begin{eqnarray}\label{eq:error_level}
	\varepsilon_E=\frac{\sigma}{\sqrt{N_g}}\;\;,
	\end{eqnarray}
	where $\sigma$ is the standard deviation.
	The standard deviation of the total particle number is given by 
	\begin{eqnarray}
	\sigma_n = \sqrt{\sum_s\frac{1}{N_{{\rm mk},s}-1} 
		\left[
		\sum_{p=1}^{N_g} w_{s,p}^2 -\frac{1}{N_{{\rm mk},s}} \left(\sum_{p=1}^{N_g} w_{s,p}\right)^2 +\varepsilon_{{\rm Var},n}
		\right] } \;\;,
	\end{eqnarray}
	where $\varepsilon_{{\rm Var},n}$ is the statistical error of the variance of the total perturbed density, which is a minor correction to the first term when the marker number is sufficiently large. 
	The standard deviation of the total current is given by
	\begin{eqnarray}
	\sigma_j = \sqrt{\sum_s\frac{q_s^2}{N_{{\rm mk},s}-1} 
		\left[
		\sum_{p=1}^{N_g} (v_{\|,p}w_{s,p})^2 -\frac{1}{N_{{\rm mk},s}} \left(\sum_{p=1}^{N_g} v_{\|,p}w_{s,p}\right)^2+\varepsilon_{{\rm Var},j}
		\right] } \;\;,
	\end{eqnarray}
	where $\varepsilon_{{\rm Var},j}$ is the statistical error of the variance of the total perturbed current. The error of the total perturbed density and current are shown in Fig.~\ref{fig:noise1d}. The standard deviation is reasonably low during the nonlinear stage, suggesting the quality of the simulation. The two cases with different numbers of markers give a similar evolution of $\sigma_n$ and $\sigma_j$, which indicates that the noise level is decreasing according to Eq.~\eqref{eq:error_level}.
	In addition, the error increases slowly but does not change dramatically in the nonlinear stage, indicating that it does not significantly worsen. 
	
\begin{figure}
		\centering
		\includegraphics[width=0.48\textwidth]{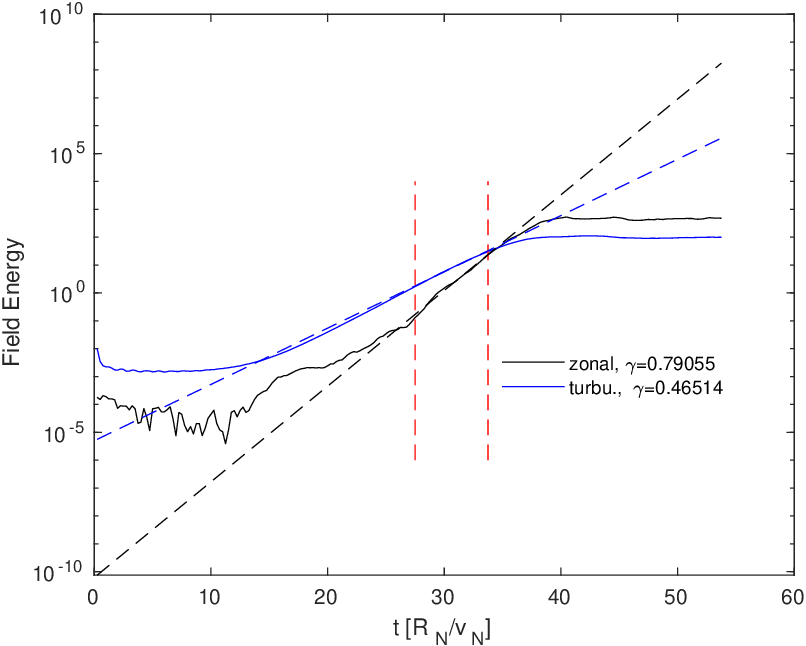}
		\caption{Time evolution of field energy for multiple $n$ simulation using 3D field-align FEM for nominal parameters ($\rho_N=0.0033422$ m) except $m_i/m_{\rm e}=100$. The growth rates $\gamma$ of the zonal component ($n=0$) and the turbulent part ($n\ne 0$)  of the total field energy are calculated using the linear fit of the time evolution of the logarithmic  total field energy during the stable linear stage indicated by the two dashed red lines.}
		\label{fig:multin_energy1d}
	\end{figure}

	The time evolution of the total field energy is shown in Fig. \ref{fig:multin_energy1d}. The zonal part ($n=0$) and the turbulent part ($n\ne0$) are separated and the corresponding total field energy is calculated. The growth rate of the zonal part is close to twice of the turbulent part, supporting that the zonal component is due to the beat-driven excitation \cite{chen2024beat,chen2024drift}. 
	
\begin{figure}
		\centering
		\includegraphics[width=0.48\textwidth]{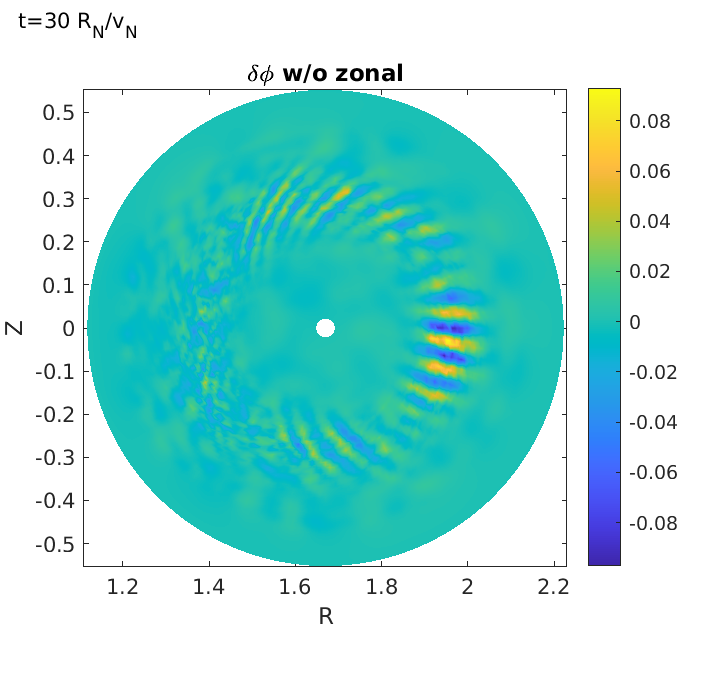}
		\includegraphics[width=0.48\textwidth]{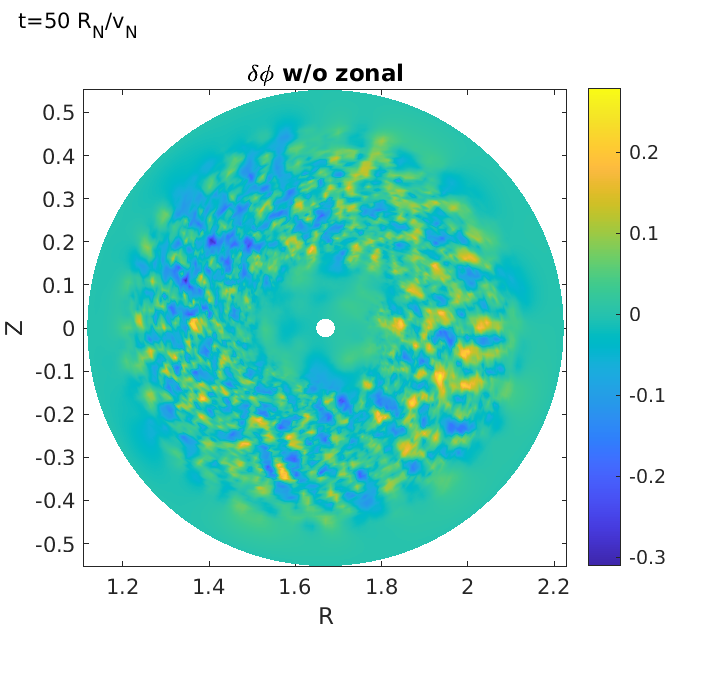}
		\caption{The 2D mode structure in the linear stage at $t=30R_N/v_N$ (left) and in the nonlinear saturated stage at $t=50R_N/v_N$ (right).}
		\label{fig:mode2d_turb_evolution}
	\end{figure}

	The turbulent part of the 2D mode structure is also visualized. The most unstable mode appears in the linear stage and becomes dominant, as shown in the left frame of Fig.~\ref{fig:mode2d_turb_evolution}. Note that the exponentially growing stage is not very long since we start the simulation with the initial perturbed level of $\delta f/f_0\sim10^{-3}$ for nonlinear cases and the 2D mode structure is not a pure state of a single toroidal harmonic but a state with several unstable toroidal harmonics. As a result, the 2D mode structure is not as pure as the single $n$ linear result in Fig.~\ref{fig:mode2d}. 
	In the nonlinear stage, the most unstable mode reaches its saturation level, and the turbulence spreads in the radial direction, as shown in the right frame where the zonal component is extracted to illustrate the features of the turbulent part. 
	
\begin{figure}
		\centering
		\includegraphics[width=0.95\textwidth]{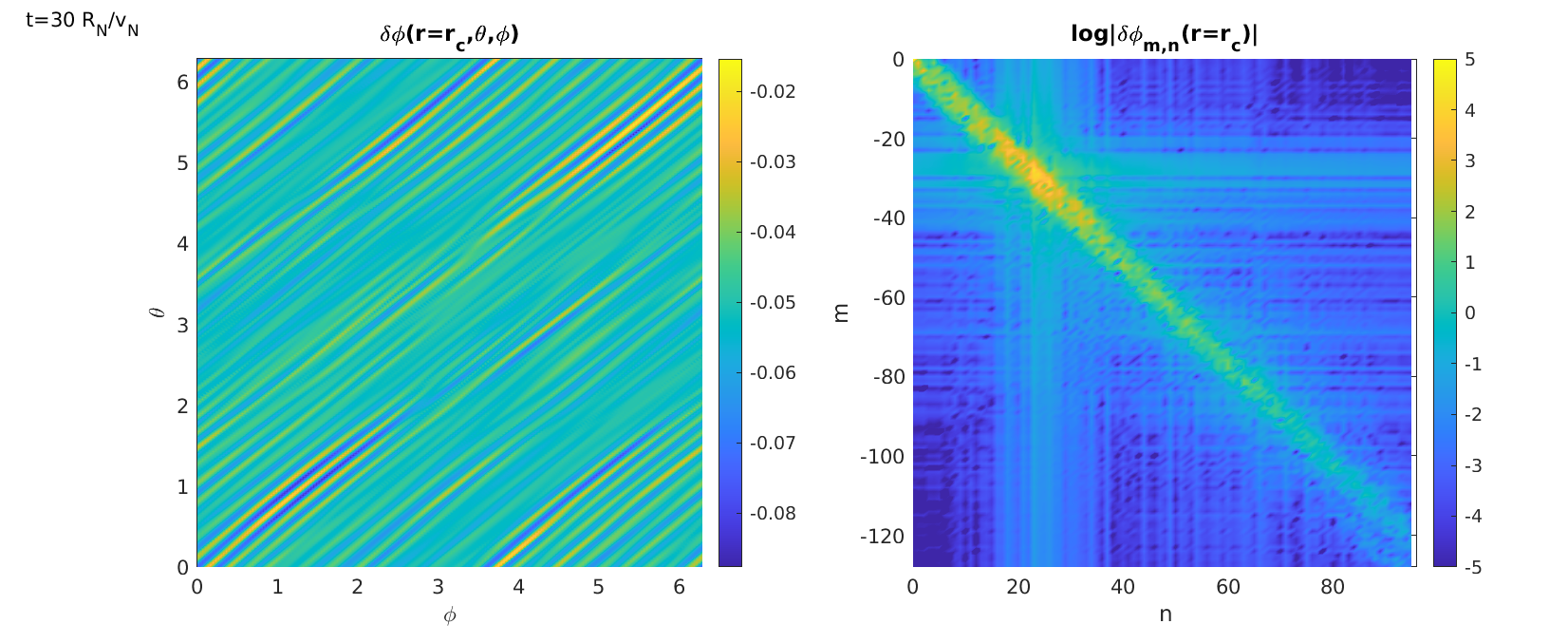}
		\includegraphics[width=0.95\textwidth]{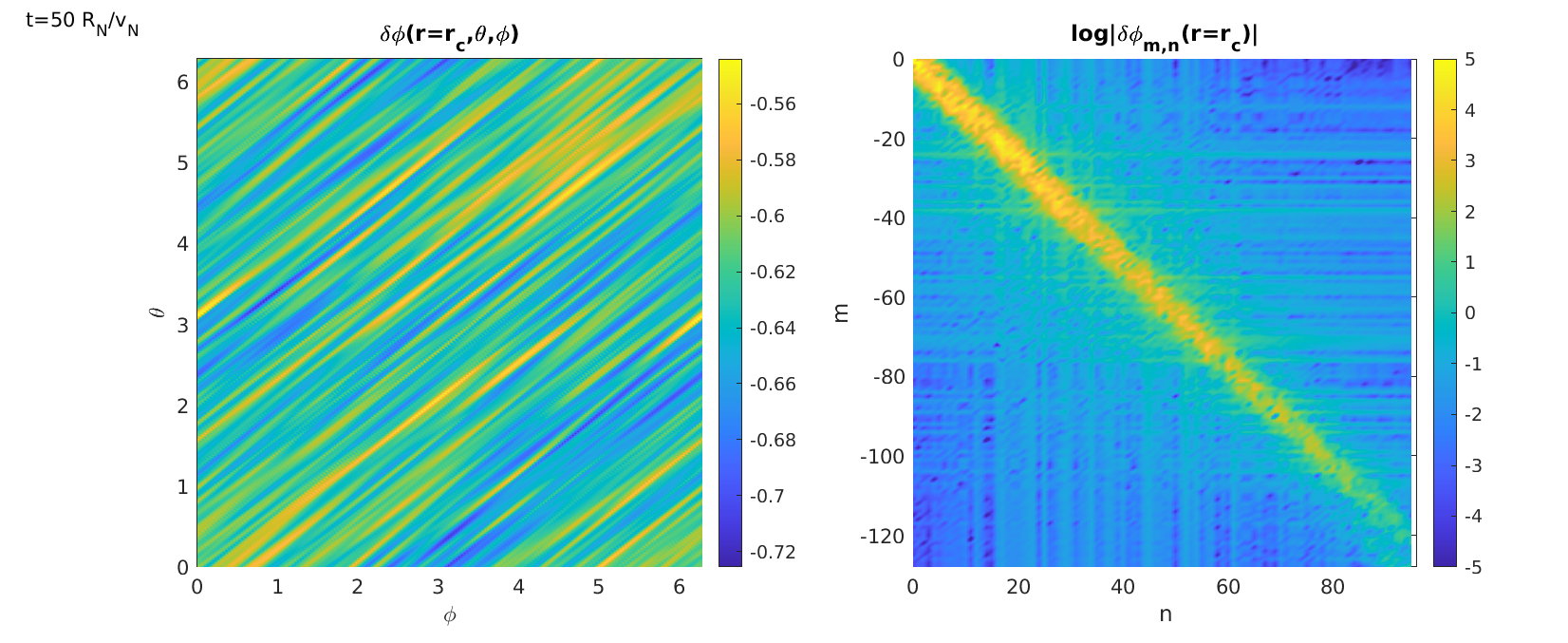}
		\caption{The mode structure $\delta\phi(r_c,\theta,\phi)$ in the magnetic flux surface at $r=r_c$ (left) and the Fourier spectrum $\delta\phi_{m,n}(r_c)$ (right). The linear and nonlinear structures are shown in the upper frame and lower frame respectively. }
		\label{fig:mode2d_turb_mn}
	\end{figure}

	The mode structure in the magnetic flux surface at $r=r_c=0.5a$ is visualized in the left frames of Fig.~\ref{fig:mode2d_turb_mn}. The mode structures are aligned along the magnetic field lines. Note although the parallel grid number is small ($N_\phi=8$), the construction of the field in the toroidal direction is possible since the parallel mode structure is smooth. The Fourier components are calculated and the logarithmic amplitude is shown in the right frames. In the linear stage  at $t=30R_N/v_N$, the peak of the spectrum is at $n\approx 25$, consistent with the linear results in Fig.~\ref{fig:cbc_nominal_singlen}. In the nonlinear stage at $t=50R_N/v_N$, other toroidal harmonics also grow. Specifically, the low $n$ harmonics have a significant amplitude, consistent with previous particle-in-cell gyrokinetic turbulence simulations \cite{wang2011trapped}. 
	
\begin{figure}
		\centering
		\includegraphics[width=0.47\textwidth]{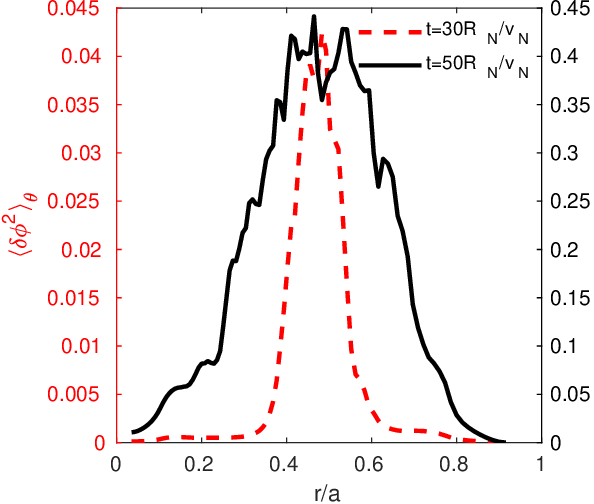}
		\includegraphics[width=0.44\textwidth]{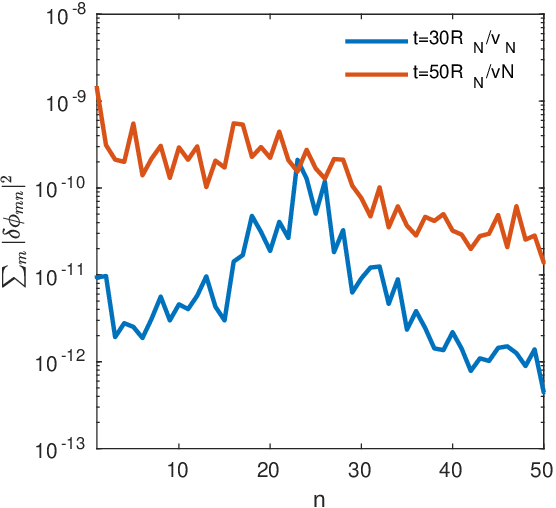}
		\caption{The radial mode structure (left) and the spectrum of the toroidal harmonics (right) in the linear stage ($t=30R_N/v_N$) and nonlinear stage ($t=50R_N/v_N$). }
		\label{fig:mode1d_turb_r_n}
	\end{figure}

	The features of the nonlinear evolution are summarized in Fig.~\ref{fig:mode1d_turb_r_n}. The radial profile of the turbulence intensity is calculated by extracting the zonal component ($n=m=0$) and integrating it along the toroidal and the poloidal directions as shown in the left frame. In the linear stage ($t=30R_N/v_N$), the mode is localized near $r=0.5a$, as indicated by the red dashed line. In the nonlinear stage ($t=50R_N/v_N$), the nonlinear spreading occurs and the radial structure of $\langle\delta\phi^2\rangle_\theta$ is broader than that of the linear structure, consistent with that in Fig.~\ref{fig:mode2d_turb_evolution} and the previous simulations \cite{mishchenko2023global}. 
	The spectrum of the toroidal harmonics $\delta\phi_n\equiv\sum_m|\delta\phi_{mn}|^2$ is calculated by summing up the poloidal harmonics (over $m$) for each $n$. Consistent with the spectrium pattern in the right frame of Fig.~\ref{fig:mode2d_turb_mn}, in the linear stage ($t=30R_N/v_N$), the spectrum of the toroidal harmonics peaks at $n\approx25$ as shown in the right frame of Fig.~\ref{fig:mode1d_turb_r_n}. In the nonlinear stage ($t=50R_N/v_N$), other toroidal harmonics especially the low $n$ harmonics also grow to a certain magnitude as indicated by the red line. Note that the simulation time in our nonlinear simulations is limited to $50R_N/v_N$  merely to demonstrate the capability of the piecewise field-aligned FEM by simulating the early nonlinear evolution. Longer simulations of the turbulent time scale \cite{bottino2011global} rely on the complete form of the gyro centers' equations of motion and the noise control schemes such as the coarse-graining algorithm~\cite{chen2007coarse} and the weight smoothing operator~\cite{sonnendrucker2015split}.
	
	\section{Conclusions}
	\label{sec:conclusions}
	We have formulated and implemented a piecewise field-aligned finite element method in tokamak geometry with magnetic shear. On one hand, the computational grids are aligned  in the traditional pattern and the centers of the basis functions are located at the grid vertices. On the other hand, the finite element basis functions are defined on piecewise field-aligned coordinates. The linear and nonlinear simulations demonstrate the features of this scheme. Our discussions and results are in the framework of the finite element method (FEM), consistent with the previous theoretical and numerical work \cite{cheng1985high,connor1979high,beer1995field,zonca2014theory,lu2012theoretical,scott2001shifted}, but extending it to FEM. The good properties of this scheme are as follows. 
	\begin{enumerate}
		\item It is formulated in the framework of the finite element method and conservation properties are inherited since the partition of unity is satisfied.
		\item Strong grid deformation is avoided due to the piecewise treatment in defining the finite element basis functions.
		\item High efficiency is expected in multi-$n$ nonlinear simulations due to the reduced grid number in the parallel direction.
		\item It applies to particle (Lagrangian) and Eulerian schemes using the finite element method.
		\item The separation of the parallel direction from the two perpendicular directions  enables the optimized treatment in the parallel direction for solving the ideal Ohm's law \cite{hatzky2019reduction}.
		\item The combination with a toroidal Fourier filter, partial torus simulations, and the application with open field lines is possible.
	\end{enumerate}
	
	The piecewise field-aligned FEM has been implemented in TRIMEG-GKX code for the electrostatic particle simulations with drift kinetic electrons and gyrokinetic ions. The Cyclone base case benchmark shows reasonable agreement between the TRIMEG-GKX results and the GENE results regarding the simplification in TRIMEG-GKX. The single toroidal harmonic simulations show agreement between the traditional 2D1F solver (FEM in the poloidal plane and the particle-in-Fourier in the toroidal direction) and the 3D field-aligned FEM solver. The multi-$n$  simulations demonstrate the capability of the field-aligned FEM in nonlinear turbulence studies. The nonlinear evolution of the ITG turbulence is simulated. The radial intensity profile and the spectrum of the Fourier modes demonstrate the nonlinear spreading of the ITG turbulence in real space and spectrum space. 
	
	As further applications, this scheme can be applied to the gyrokinetic simulations using unstructured meshes \cite{lu2024gyrokinetic,lu2019development} for electromagnetic particle simulations \cite{lu2023full,lanti2020orb5,hatzky2019reduction,mishchenko2023global}, for the whole device modeling. It can also be useful for particle simulations in stellarators \cite{kleiber2024euterpe}, especially for multi-$n$ nonlinear simulations.

	\section*{Acknowledgement}
	Z.X. Lu appreciates the inputs from Peiyou Jiang, Ralf Kleiber, Alberto Bottino, Weixing Wang and Laurent Villard. The simulations in this work were run on the TOK cluster, the Raven supercomputer, the Viper supercomputer at MPCDF, and  the MARCONI supercomputers at CINECA. 
	The Eurofusion projects TSVV-8, ACH-MPG, TSVV-10, and ATEP are acknowledged. 
	This work has been carried out within the framework of the EUROfusion Consortium, funded
	by the European Union via the Euratom Research and Training Programme (Grant Agreement No
	101052200—EUROfusion). Views and opinions expressed are however those of the author(s)
	only and do not necessarily reflect those of the European Union or the European Commission.
	Neither the European Union nor the European Commission can be held responsible for them.
	
	\newpage
	\providecommand{\newblock}{}

\end{document}